\documentclass[preprints,galaxies,review,accept,moreauthors,pdftex,10pt,a4paper]{Definitions/mdpi} 
\usepackage{hyperref}

\newcommand{\lum}{erg s$^{-1}$}
\newcommand{\Lx}{$L_{\rm X}$}
\newcommand{\Msun}{$M_{\odot}$}

\newcommand{\kms}{km s$^{-1}$}
\newcommand{\spitzer}{\textit{Spitzer}}

\firstpage{1} 
\pubvolume{xx}
\issuenum{xx}
\articlenumber{xx}
\pubyear{2020}
\copyrightyear{2020}
\history{Received: 2 November 2019; Accepted: 10 February 2020; Published: date}

\pdfoutput=1

\usepackage{multicol}
\usepackage{natbib}

\Title{From SN~2010da to NGC~300 ULX-1: Ten Years of Observations of an Unusual High Mass X-ray Binary in NGC~300}

\Author{Breanna A. Binder$^{1*}$, Stefania Carpano$^2$, Marianne Heida$^3$ \& Ryan Lau$^{4}$}

\AuthorNames{Binder et al.}

\address{%
$^{1}$ \quad Department of Physics \& Astronomy, California State Polytechnic University, Pomona, 3801 W. Temple Ave, Pomona, CA, USA 91768 \\
$^{2}$ \quad Max Planck Institute for Extraterrestrial Physics, Giessenbachstra{\ss}e 1, 85748 Garching, Germany\\
$^{3}$ \quad Division of Physics, Mathematics \& Astronomy, California Institute of Technology, 1200 E. California Blvd, Pasadena, CA, USA 91125\\
$^{4}$ \quad Department of Space Astronomy \& Astrophysics, Japan Aerospace Exploration Agency, 3-1-1 Yoshinodai, Chuo-ku, Sagamihara, Kanagawa 252-5210, Japan}

\corres{Correspondence: babinder@cpp.edu; Tel.: +1-909-869-4029}

\abstract{In 2010 May, an intermediate luminosity optical transient was discovered in the nearby galaxy NGC~300 by a South African amateur astronomer. In the decade since its discovery, multi-wavelength observations of the misnamed ``SN 2010da'' have continually re-shaped our understanding of this high mass X-ray binary system. In this review, we present an overview of the multi-wavelength observations and attempts to understand the 2010 transient event and, later, the re-classification of this system as NGC~300 ULX-1: a red supergiant + neutron star ultraluminous X-ray source. }


\begin{document}

\section{Initial Discovery}
\begin{figure}
\centering
\begin{tabular}{cc}
     \includegraphics[height=0.45\linewidth]{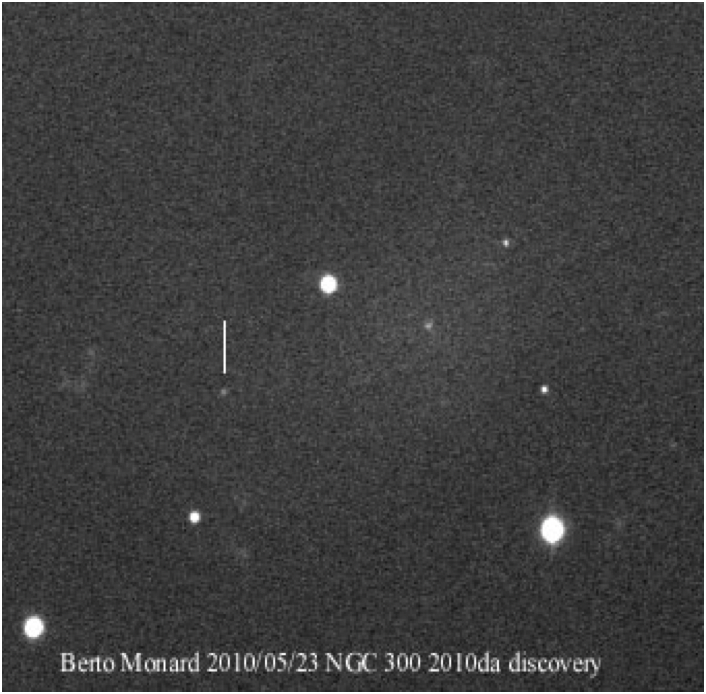} &
     \includegraphics[height=0.45\linewidth]{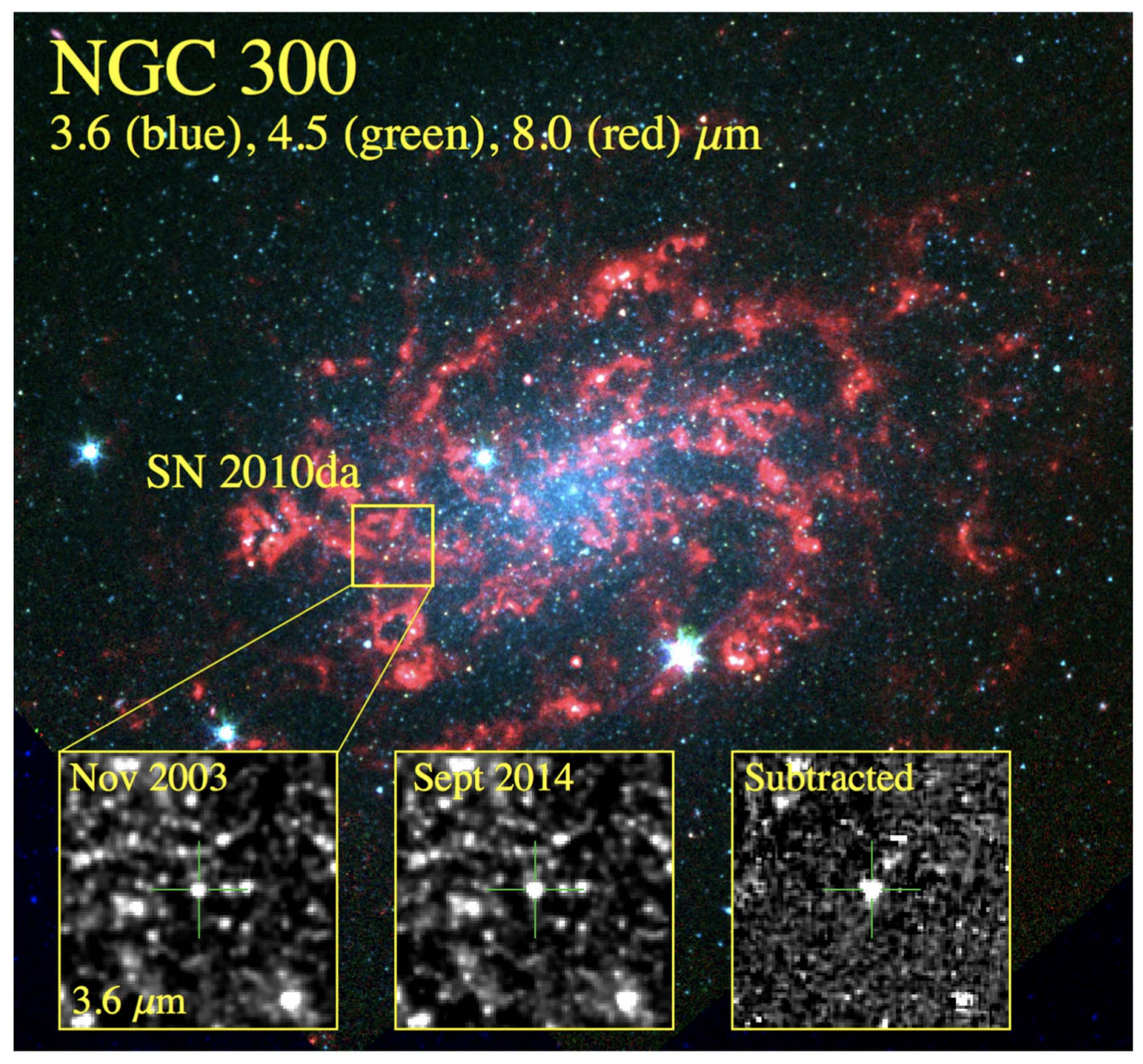} \\
\end{tabular}
\caption{{\it Left}: the initial discovery image \citep{GalYam+13} of SN~2010da by L. A. G. Monard \citep{Monard10}. {\it Right}: an RGB-rendered IR composite image of NGC~300, with insets showing the 3.6 $\mu$m image of the progenitor ($\sim$6.5 years pre-outburst), $\sim$4.5 years post-outburst, and the difference subtracted image. Reproduced with permission from \citet[][their Figure 1]{Lau+16}.}
\label{figure:discovery}
\end{figure}

On 2010 May 24, the amateur astronomer L. A. G. Monard of Pretoria, South Africa, reported the discovery of an optical transient in the nearby spiral galaxy NGC~300, shown in Figure~\ref{figure:discovery} \citep{Monard10}. The event was given the supernova (SN) designation SN~2010da. The apparent $V$-band magnitude from the discovery images (16.0$\pm$0.2 mag), however, immediately presented a contradiction: at the distance of the host galaxy (2 Mpc \citep{Dalcanton+09}), the implied absolute magnitude ($M_{\rm V}\sim -10.5$ at maximum; \citep{Khan+10atel}) was too low to be consistent with a true SN explosion. Follow-up spectroscopy confirmed that the transient event was not a genuine SN, and instead resembled a luminous blue variable (LBV)-like outburst from a dust-enshrouded massive star \citep{Chornock+10, EliasRosa+10}. Multi-color imaging of the outburst was obtained with the SMARTS 1.3m telescope at Cerro Tololo $\sim$2 days after the initial discovery \citep{Bond10} and $\sim$3 weeks post-outburst \citep{Prieto+10}.

SN~2010da thus joined the heterogeneous group of objects known as ``supernova impostors,'' which are frequently interpreted as massive stars experiencing non-terminal eruptions or outbursts \citep{Bond10, EliasRosa+10}. Additional spectroscopy from the SOAR 4.1m telescope spanned 400-860 nm \citep{EliasRosa+10} showed similarity to the SN impostor SN~1997bs \citep{VanDyk+00}. The blue continuum emission of SN~2010da shortly after outburst was dominated by strong H$\alpha$ emission, with a full width at half maximum (FWHM) of $\sim$1000 \kms\ and no evidence of a P Cygni profile, along with other narrow Balmer lines and a possible detection of weak He II $\lambda$4686 \citep{EliasRosa+10}. These results stood in contrast to those obtained with near-simultaneous GMOS on Gemini-South by \citet{Chornock+10}, which observed an H$\alpha$ FWHM of $\sim$660 \kms\ and higher-order Balmer lines with clear P-Cygni profiles. These spectral properties led \citet{Chornock+10} to speculate that the SN~2010da event originated in ``an LBV-like outburst of a massive star.''

Although it was immediately acknowledged that the outburst luminosity of SN~2010da was too low to be a true SN, it was \textit{also} too low to be an LBV eruption; the 2010 outburst of SN~2010da was $\sim$200 times fainter than the Great Eruption of $\eta$ Car in 1845.  To further complicate the interpretation, a bright X-ray point source was detected by the {\em Neil Gehrels Swift Observatory} within hours of the optical detection, with an approximate 0.2-10 keV luminosity of $(6.0\pm0.6)\times10^{38}$ \lum\ \citep{Immler+10}. The X-ray emission was initially attributed to stellar processes (the initial choice of X-ray spectral model by \citet{Immler+10} was a thermal plasma, the usual choice for erupting stars and SN impostors, and \citet{Smith+11} describe the high $L_X$ of SN~2010da in the context of shocks during giant stellar eruptions), however the brightest known stellar X-ray emitters (colliding wind binaries) have luminosities on the order of \Lx$\sim$10$^{35}$ \lum\ \citep{Guerrero+08} and the highest observed X-ray luminosities from an LBV outburst reach only $\sim10^{34}$ \lum\ \citep{Naze+12}, orders of magnitude lower than what was observed by {\em Swift}. A new interpretation was needed to explain the high observed X-ray emission.

\subsection{Understanding the SN~2010da System Prior to Outburst}
\begin{figure}
    \centering
    \includegraphics[width=0.5\linewidth]{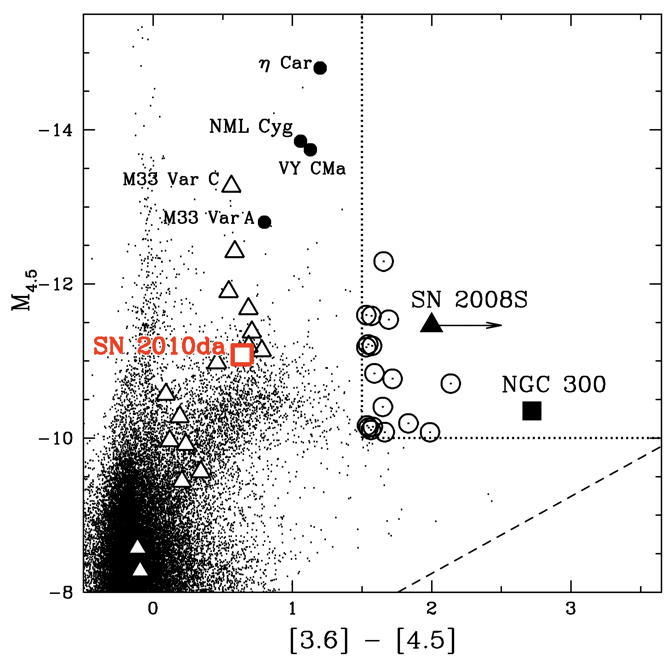}
    \caption{The mid-IR CMD showing the SN~2010da progenitor (red square) and local LBVs and LBV candidates (filled circles and triangles, respectively). Open circles and filled triangles show the CMD locations of LBV-like SN ``impostors.'' Reproduced with permission from \citet{Khan+10atel}.}
    \label{figure:progenitor_IR_CMD}
\end{figure}

As photometric and spectroscopic follow-up of SN~2010da commenced, several groups searched for the SN~2010da ``progenitor'' system (e.g., the system prior to the 2010 transient event) in archival observations. Mid-infrared (IR) photometry from archival {\em Spitzer}/IRAC imaging data obtained in 2007 December \citep{Khan+10atel, Laskar+10} revealed that the progenitor of SN 2010da was heavily obscured and shared a similar position in a mid-IR color–magnitude diagram (CMD; see Figure~\ref{figure:progenitor_IR_CMD}) as luminous blue variable (LBV) candidates \citep{Massey+07, Khan+10, Thompson+09, Laskar+10}.

\begin{figure}
    \centering
    \includegraphics[width=0.65\linewidth]{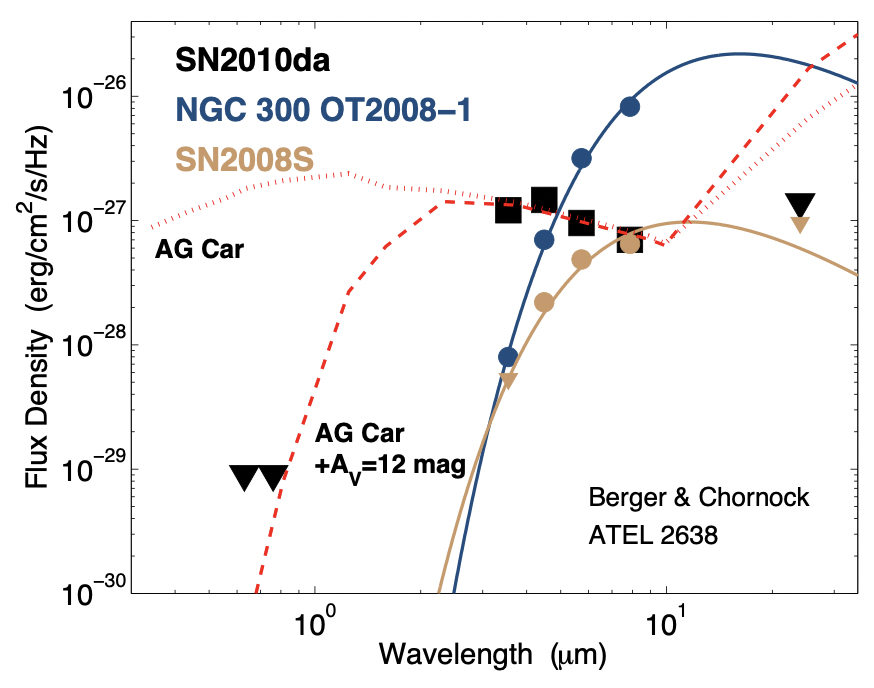}
    \caption{The IR SED of SN~2010da progenitor (black squares show measurements, black downward facing triangles show upper limits). The SEDs of two other optical progenitors, NGC~300 OT2008-1 and SN~2008S, are shown (in blue and brown, respectively) for comparison. The dotted red line shows the SED of the known LBV AG Car; the addition of 12 mag extinction is required for this SED to fit the observed fluxes of SN~2010da (dashed red line). Reproduced with permission from \citet{Berger+10}.}
    \label{figure:progenitor_IR_SED}
\end{figure}

The SN~2010da progenitor was found to have brightened by $\sim$0.5 mag in the $\sim$6 months prior to the 2010 outburst, indicating a period of enhanced activity prior to the optical outburst \cite{Laskar+10}. Although IR observations of the progenitor system showed consistency with known Local Volume LBVs \cite{Laskar+10,Khan+10atel}, archival optical and 24 $\mu$m images provided a hint at a discrepancy. \citet{Berger+10} reported the non-detection of the SN~2010da progenitor in archival Magellan/Megacam images, which were obtained approximately one year earlier to follow up a prior transient in NGC~300 \citep{Berger+09,Bond+09}. No optical progenitor was detected in either the $r$ or $i$ band to a limiting magnitude of 24 mag. A first look at the progenitor IR spectral energy density (SED, shown in Figure~\ref{figure:progenitor_IR_SED}) of SN~2010da showed it to be significantly bluer than previously studied optical transients, as well as the known Galactic LBV AG Car \citep{Berger+10}. It was therefore proposed that the SN~2010da progenitor was enshrouded by a thick cocoon of dust ($A_{\rm V}>12$ mag), most of which was destroyed in the subsequent outburst. Although this hypothesis accommodated the optical upper limits, \citet{Berger+10} noted that the 24 $\mu$m flux limit was still roughly ``an order of magnitude lower than expected based on the SED of AG Car''.

\section{SN~2010da as a High Mass X-ray Binary}
Four months after the initial outburst, NGC~300 was observed with the {\em Chandra X-ray Observatory}, and SN~2010da was observed with a 0.35-8 keV X-ray luminosity of $\sim4\times10^{37}$ \lum\ \citep{Binder+11}. In order to reconcile the X-ray observations with the mid-IR and optical studies, \citet{Binder+11} suggested that SN~2010da was actually a high-mass X-ray binary (HMXB) in NGC~300, possibly composed of a neutron star (NS, due to the hardness of the X-ray spectrum) and some sort of evolved, massive (``LBV-like'') companion. This observation revealed the first instance of a SN impostor (or intermediate luminosity optical transient) originating in an HMXB.

\begin{figure}
    \centering
    \begin{tabular}{cc}
        \includegraphics[height=0.35\linewidth,clip=true,trim=2.1cm 2.1cm 0 0]{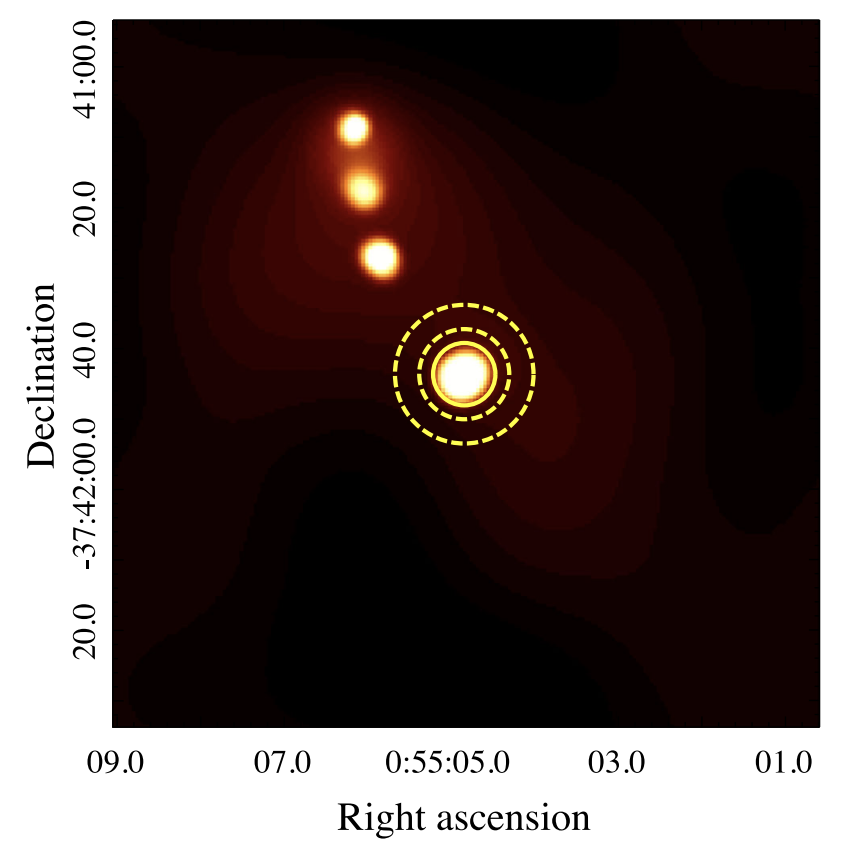} &
        \includegraphics[height=0.35\linewidth]{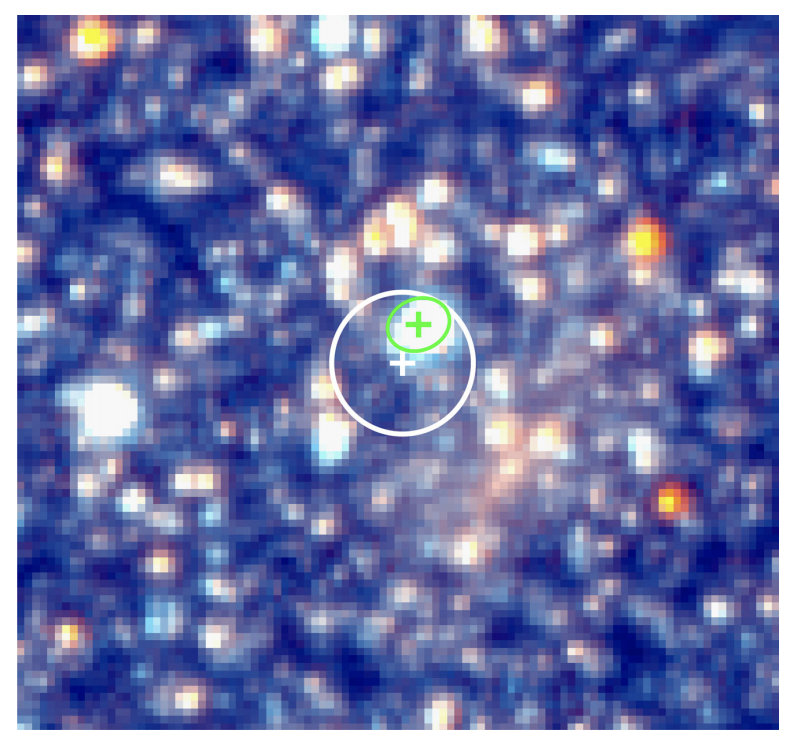} \\
    \end{tabular}
    \caption{{\it Left}: Adapted from Figure~1 in \citet{Binder+16}, showing the {\em Chandra} image of SN~2010da. The yellow circle indicates the location of the X-ray source coincident with SN~2010da (the three fainter sources to the upper-left are unrelated). {\it Right}: Figure~2 from \citet{Binder+16} showing the {\em Hubble}/ACS image of the same region. The white cross and circle indicate the location of the X-ray source; the green cross and circle show the location of the likely massive donor star.}
    \label{figure:Binder16_images}
\end{figure}

In 2016, a trio of papers investigating the nature of SN~2010da under the HMXB paradigm were published. New {\em Chandra} and {\em Hubble Space Telescope} imaging obtained in 2014 \citep{Binder+16} revealed a variable X-ray source coincident with the spatial position of SN~2010da and a bright, blue optical source consistent with a blue supergiant or LBV-like star (shown in Figure~\ref{figure:Binder16_images}); the prevalence of young stars in the vicinity of SN~2010da suggests the system has an age of $\leq$ 5 Myr from population synthesis modeling. \citet{Lau+16} presented detailed IR SED modeling of the SN~2010da progenitor and progeny, which indicated that a large fraction of the dust that initially enshrouded the SN~2010da progenitor was destroyed during the 2010 outburst. The relatively blue color of the 2010 outburst compared to other SN impostors was noted by \citet{Bond10}, and low reddening in 2014 ($A_{\rm V}\sim0.4$ mag) was also inferred from stellar population synthesis modeling in \citet{Binder+16}. Using the radiative tranfer code DUSTY \cite{Ivezic+97}, \citet{Lau+16} determined the inner radius of dust emission from the SN~2010da system to be $\sim$20-40 AU at $\sim$2.5 years post-outburst. The close proximity to the central system (shock speeds of $\sim$660 \kms\ from the outburst H$\alpha$ emission line FWHM \cite{Chornock+10}, would sweep surviving dust out to $\sim$300 AU assuming the stellar wind/dust interaction models of \citep{Kochanek11}) suggests that the ``post-outburst emission in the mid-IR is primarily associated with newly formed dust from the surviving central system.''

High and medium/low resolution optical spectroscopy with Gemini and multi-wavelength photometry of the pre- and post-outburst SN~2010da system was described in \citet{Villar+16}; notably, multi-component emission lines of hydrogen Balmer, Paschen, and Ca~II revealed a complex circumstellar environment, although the continuum emission was unabsorbed. The presence of He~II $\lambda$4686 and coronal Fe emission lines was also highlighted and first attributed to a compact companion.

It was additionally noted during this time that the X-ray energetics favored a highly eccentric binary orbit ($e>0.82$; \cite{Lau+16}), consistent with the young age inferred from the spectral population synthesis modeling of young stars in the vicinity of the SN~2010da system \citep{Binder+16}. Although mid-IR observations of the progenitor yielded firm photometric detections, no X-ray source was detected in pre-2010 archival {\em XMM-Newton} observations down to a limiting luminosity of $L_{\rm X}<(3-9)\times10^{36}$ \lum\ \citep{Binder+11}. This lack of a previous X-ray detection and young apparent age led \citet{Binder+16} to hypothesize that the 2010 outburst may have represented the first onset of X-ray production in the system, consistent with simulations of HMXB formation \cite{Linden+10}.

\section{The Ultraluminous X-ray Source}
In 2016, the SN~2010da HMXB system was serendipitously observed for the first time in a ultraluminous (ULX) state during a long and simultaneous {\em XMM-Newton}/{\em NuSTAR} observation performed between December 16 and 20 \citep{Carpano+18}. The X-ray flux of the object increased by roughly two orders of magnitude above the 2014 {\em Chandra} observations, yielding a 0.3--30\,keV unabsorbed luminosity of $\sim4\times10^{39}$ \lum. Off-nuclear point sources exhibiting such high X-ray luminosities (e.g., exceeding the Eddington limit of an isotropically-emitting 10 \Msun\ black hole) are collectively referred to as ultraluminous X-ray sources \citep[ULXs; ][]{Kaaret+17}. The system hence received a new designation: NGC~300 ULX-1.

ULXs were initially thought to be powered by intermediate-mass black holes ($M_{\rm BH}\sim100-1000$ \Msun). However, it is now known that the majority of these objects are otherwise normal X-ray binaries with stellar-mass compact objects accreting at super-Eddington rates or with non-isotropic emission geometries \citep{Kaaret+17}. Models and simulations of super-Eddington accretion around black holes \citep{SS73,Poutanen+07,Narayan+17} predict a geometrically and optically thick accretion disk around the compact object, with an optically thin, evacuated funnel forming along the object's axis of rotation. Excess radiation pressure is converted into mechanical energy, which launches material away from the compact object in a wide, ultrafast outflow (UFO, with speeds of $\sim0.1-0.3c$). The accretion disks around NSs are more complex due to the presence of strong, persistent magnetic fields, which truncate the inner edge of the accretion disk at much larger radii than for black holes (the magnetospheric radius, $R_{\rm M}$, is often used as a proxy for the inner disk radius around a magnetized neutron star; \citep{Vasilopoulos+19,Christodoulou+18}). Unlike in their BH counterparts, an accretion column may form in a NS-ULX system, and is thought to be the source of the hard X-ray emission observed from these objects \citep{Mushtukov+17,Walton+18b}. Strong winds from NS-ULXs are therefore indicative of super-Eddington accretion flows beyond $R_{\rm M}$ \citep{Walton+18b}.

\begin{figure*}
    \centering
    \includegraphics[width=0.55\linewidth]{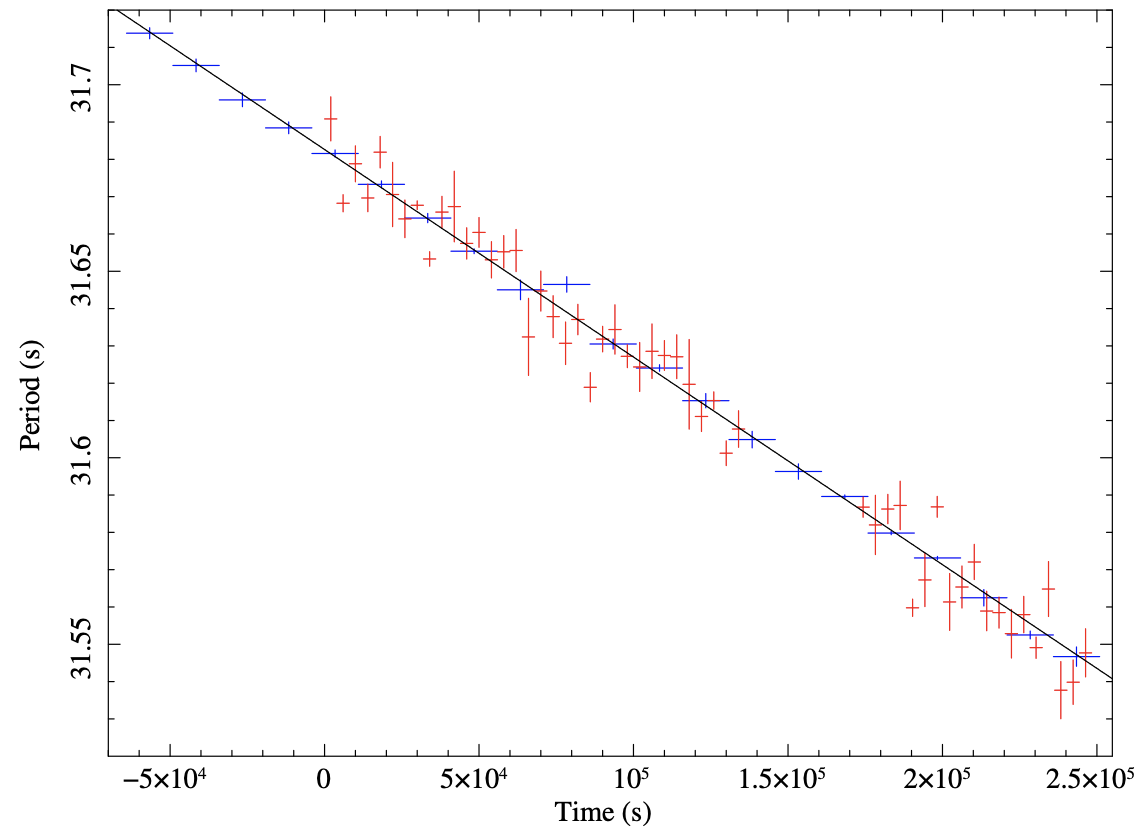}
    \caption{Figure~1 from \citet{Carpano+18}, showing the period evolution during the deep {\em XMM-Newton}/{\em NuSTAR} observations.}
    \label{figure:Carpano_period}
\end{figure*}

The deep {\em XMM-Newton}+{\em NuSTAR} observations of NGC~300 ULX-1 yielded numerous insights into the nature of the system, which we discuss in further detail below. Pulsations were found with an average value of 31.6\,s, confirming the compact object's identity as a neutron star (NS), and with an incredibly high spin up rate of $-5.56\times10^7 \textrm{s s}^{-1}$ \citep[][see Figure~\ref{figure:Carpano_period}]{Carpano+18}. The pulsed fraction was also extremely large, varying from $\sim$50\% at $\sim$1\,keV to $\sim$75\% in the {\em NuSTAR} band ($\sim10-20$\,keV). Evidence was also found for an ultrafast outflow (UFO) and potential cyclotron resonant scattering feature (CRSF), providing clues to the accretion geometry and interaction with the magnetic field of the NS. At the time of its discovery as a ULX, NGC~300 ULX-1 was only the fourth confirmed NS-ULX.

\subsection{Period Evolution}
Using archival data, it was shown by \citet{Vasilopoulos+18} and \citet{Vasilopoulos+19} that the spin period of NGC~300 ULX-1 was as large as 126.3\,s in a {\em Chandra} observation from 2014 November. Assuming almost steady accretion rate $\dot{M}$ after the initial outburst, the spin-period evolution could only be explained by an accretion disk formed rotating retrograde with respect to the spin of the NS, with the NS spinning down until its rotation was stopped and then starting in the opposite direction \cite{Vasilopoulos+18}. On the other hand, if $\dot{M}$ was two orders of magnitude lower in 2010 than after 2015, then there could be no spin reversal. Although caution must be used in inferring the accretion history of NGC~300 ULX-1 from 2010-2014 due to sparse X-ray observations, the changing $\dot{M}$ scenario appears unlikely, as all available observations of the observed spin evolution and nearly constant X-ray luminosity are consistent with a roughly constant mass accretion rate since the initial 2010 outburst event \citep{Vasilopoulos+19}.

A deep analysis of the pulsed period evolution from 2018 February 6 through 2018 October 20 was performed by \citet{Ray+19} using {\em NICER} data. Three spin-down glitches of magnitudes $\Delta\nu=-23, -30$ and $-43\ \mu$Hz (fractional amplitudes $\Delta\nu/\nu=-4.4, -5.5$ and $-7.7\times 10^{-4}$) were discovered over a span period of 116\,d using the coherent timing analysis technique. These glitch amplitudes are larger (and opposite in sign) than any reported radio pulsar glitch. The spin-down nature of these glitches is of particular interest since many hundreds of radio, X-ray and $\gamma$-ray pulsars have glitches that are spinning up ($\Delta\nu$>0, \citep{Ray+19} and reference therein). The authors believe that, like in the case of non-accreting pulsars, the glitches are likely resulting from a superfluid in the pulsar's inner crust and possibly core. An accurate measurement of the pulse period evolution is also generally performed to search for the orbital period of the system. The orbital movement should be revealed through a cyclic variation in the long-term pulse period trend, caused by Doppler shifts. However, to date no evidence of an orbital period has been found for NGC~300 ULX-1 in any existing data set, implying that the system is either almost face-on or that the period is longer than a year.

\subsection{The X-ray Spectrum}
\begin{figure*}
    \centering
    \includegraphics[width=0.55\linewidth]{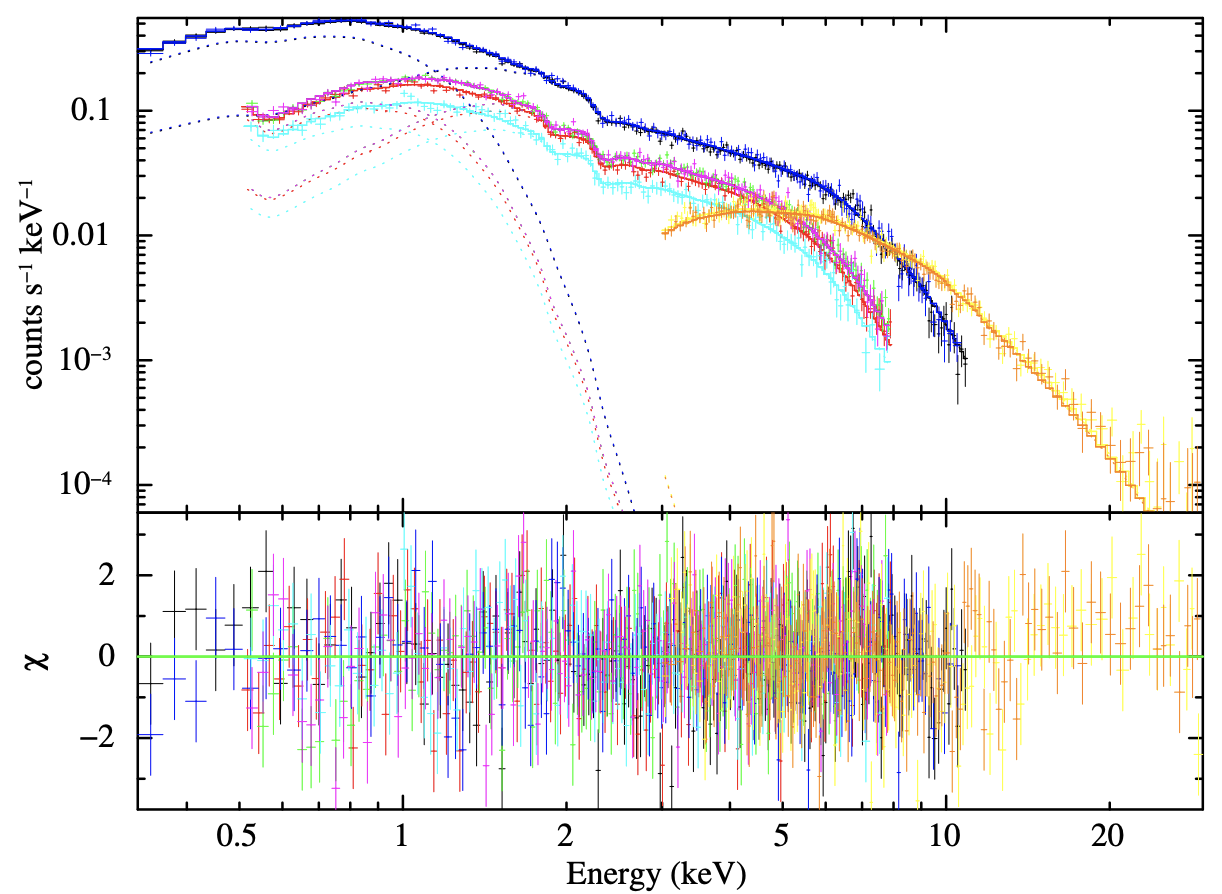} 
    \caption{The {\em XMM-Newton} and {\em NuSTAR} spectra of NGC~300 ULX-1 (top panel of Figure~3 in \citet{Carpano+18}). {\em XMM-Newton}/EPIC pn spectra are shown in black and blue, while the MOS spectra are shown in red, green, magenta and cyan. {\em NuSTAR} FPMA and FPMB spectra are shown in yellow and orange, respectively.}
    \label{figure:Carpano_spectrum}
\end{figure*}

Several detailed spectral models have been tested on both the pulsed and phase-averaged X-ray continuum of NGC~300 ULX-1 using the simultaneous {\em XMM-Newton} and {\em NuSTAR} observations from 2016 December (see Figure~\ref{figure:Carpano_spectrum}). These models attempt to describe two possible models for X-ray emission in the system. In the first model, the pulsating hard component (typically assumed to be a power law with a high-energy ``roll-off'') is thought to originate from an accretion column, while a multicolor disk blackbody represents the truncated inner accretion disk. This model was developed to describe observations of several nearby accreting pulsars \citep{Becker+07}. Alternatively, the characteristic X-ray ``roll-off'' can be described with a second hot thermal component ($E\geq2$ keV) that originates from an optically thick accretion envelope that engulfs the NS \citep{Mushtukov+17}. Pulsations in the hard component are then interpreted as a significant temperature gradient in the spinning accretion envelope as material falls from the magnetospheric radius to the base of the envelope at the top of the accretion column.

In \citet{Carpano+18}, the authors successfully fit a model composed of two absorption components (one for the interstellar absorption in the light of sight, and one partial-covering absorber in the vicinity of the neutron star), with a disk blackbody model (representing the inner edge of the accretion disk) and a power law component (to account for harder radiation from near the NS) with high energy cut-off. This model could describe the outburst {\em XMM-Newton} spectrum from 2010 May as well, with the addition of a Fe-K$\alpha$ fluorescence line at 6.4\,keV and more significant local absorption than in 2016. This model was used to infer an unabsorbed 0.3--30 keV luminosity of 4.7$\times10^{39}$ \lum\ and firmly establish this source as a pulsar-ULX.

\citet{Kosec+18} used a similar model for the X-ray continuum using both the soft X-ray spectra from the {\em XMM-Newton RGS} instrument and in the harder band from the {\em XMM-Newton} EPIC (pn, MOS1 and MOS2) instruments, as well as {\em NuSTAR}, combining a hard power law component with a multicolor blackbody; their detailed analysis of fit residuals shows evidence of a UFO at $>$3$\sigma$ significance from the NS. If confirmed, this would be the first direct evidence of a UFO (with a projected velocity $\sim0.22c$, similar to other ULXs) from a pulsar-ULX, consistent with the theoretical predictions of \citet{Mushtukov+19}. The UFO was detected in only one of two {\em XMM-Newton} observations; the lack of  detection in the other {\em XMM-Newton} observation may be explained by the presence of clumps in the absorber or a slight change in viewing angle of the accretion flow.

More detailed and physically-motivated studies of the broadband X-ray continuum were performed by \citet{Walton+18} and \citet{Koliopanos+19}. In \citet{Walton+18}, the pulsed spectrum of NGC~300 ULX-1 was constructed by extracting the data from the brightest and faintest quarters of the pulsed cycle and removing the later from the former ("pulse-on"--"pulse-off"; the pulse profile is nearly sinusoidal, \citep{Carpano+18}). The spectrum was modeled as a power-law with a high-energy cut-off (e.g., assuming the accretion column scenario for the X-ray emission) and a high-energy excess was observed in the fit residuals. The addition of a \texttt{SIMPL} component to the model -- Comptonization in which a fraction of the photons in an input seed spectrum are scattered into a power-law component -- significantly improved the fit \cite{Steiner+09}. However, the residuals were also explained by the existence of a broad cyclotron resonant scattering feature (CRSF) causing a deficit in the $\sim$5--20\,keV band. CRSFs are due to resonant scattering of X-ray photons with electrons embedded in the magnetic field and, when detected,allow a direct estimation of the magnetic field intensity: $E_{\rm CRSF}=11.57\times B_{12}(1+z_{\rm grav})$, where $E_{\rm CRSF}$ is the energy of the CRSF absorption line center, $B_{12}$ is the magnetic field strength (in units of $10^{12}$ G), and $z_{\rm grav}$ is the gravitational redshift of the line forming region ($z_{\rm grav}\leq0.25$, since the line must originate at or beyond the surface of the NS). A Gaussian absorption line at $E\sim$ 12.8\,keV improved the fit significantly with respect to a simple single-component model, and implies the existence of a magnetic field with $B\sim10^{12}$\,G close to the surface of the NS. This value is very close to what was measured in \citet{Carpano+18} from the spin-up rate, and similar to the magnetic field strengths of Galactic X-ray pulsars \citep{Caballero+12}.

\citet{Koliopanos+19} investigated whether the phase-averaged spectrum could be described as a multicolor thermal emission from an optically thick accretion envelope (e.g., the second model described above). Using a Bayesian X-ray analysis, they found that a multicolor accretion envelope model without a Gaussian absorption feature was preferred over the accretion column model. The soft, $\sim$0.3 keV multicolor disk component in their model implies an inner accretion disk radius of 400-800 km (assuming an inclination angle of 60$^{\circ}$, and is consistent with a geometrically thin accretion disk truncated at $R_{\rm M}$ (assuming an equatorial $B$-field strength of $\sim3-7\times10^{12}$ G). In the accretion column model, both the flux and the slope of the high-energy power-law component are expected to change. This could be due to a change in  the inclination angle at which the observer views the column during the ``pulse-on'' and ``pulse-off'' phases; during the ``pulse-on'' phase, the accretion column is closer to face-on, which increases the scattering optical depth and hardens the power-law slope \citep{Koliopanos+19,Mushtukov+17}. This behavior is not observed in NGC~300 ULX-1, where the fitted ``pulse-on'' power-law slope is marginally softer than in the ``pulse-off'' spectrum \citep{Koliopanos+19}.

The presence of a CRSF and the shape of the X-ray continuum in NGC~300~ULX-1 is therefore still a matter of debate. Further observations by next-generation, hard imaging and spectroscopic telescopes are needed to measure the shape of the NGC~300 ULX-1 spectrum above $\sim$40 keV, where the two models begin to significantly diverge.

\subsection{Beaming}
In addition to the deep {\em XMM-Newton} and {\em NuSTAR} observations, {\em Swift}/XRT obtained shorter exposures of NGC~300 ULX-1 beginning in 2016 April. During the summer of 2017, \citet{Binder+18} obtained four nights (spread over a $\sim$6 week period) of Gemini optical spectroscopy that were nearly simultaneous with {\em Swift} observations. As was found by \citet{Villar+16}, strong multi-component emission features were detected (the FWHM of H$\alpha$ was found to be $\sim$300 \kms, roughly half that during outburst). The simple continuum model of \citet{Carpano+18} was used to estimate the unabsorbed X-ray luminosity in each {\em Swift} observation, as the low number of counts and restricted energy range of each observation prohibited more detailed continuum modeling. Variations between the eleven spectra could be modeled as changes in the partial covering fraction in each observation, and was interpreted as optically thick clumps in the outflow moving in and out of the line of sight to the observer. Under this model, in the observed changes in the H$\alpha$ line profile (which were also observed at earlier epochs by \citet{Villar+16}) could also be explained as changes in the orientation of the outflow.

The simultaneity of the X-ray and optical observations enabled the use of the He~II $\lambda$4686 emission feature (which is sensitive to 54-200 eV photons) as a ``photon counter'' to determine that geometric beaming did not play a significant role in boosting the observed X-ray luminosity of the system. This observation runs counter to some theoretical predictions which require a high degree of beaming to produce ULX-luminosities \citep{King+19,King09}, and some degree of beaming is needed to explain the observed X-ray pulsations. However, the sinusoidal pulse profile observed in NGC~300 ULX-1 (as well as other ULX pulsars) is inconsistent with a narrow beam (unless the observer is serendipitously observing all ULX pulsars at favorable inclinations, \citep[][and references therein]{Kaaret+17}). Using accretion torque theory and the measured spin period evolution of the NS, \citet{Vasilopoulos+19} independently concluded that, if outflows were indeed present in the system, ``no beaming is needed to explain the spectral and temporal properties'' of NGC~300 ULX-1. The He~II $\lambda$4686 emission feature is sensitive only to {\it soft} X-rays; e.g., those originating from the accretion disk primarily beyond the magnetospheric radius. In \citet{Binder+18}, $<$20\% of the total unabsorbed luminosity of NGC~300 ULX-1 was observed below 1 keV. Although the hard (and pulsed) emission from the accretion column or accretion envelope may be beamed, this emission does not significantly contribute to the He~II line luminosity. Other ULXs for which the extended He II emission nebulae can be resolved (e.g., Holmberg~II X-1, \citep{Kaaret+04}) similarly show that the He~II photoionizing emission from the ULX is, at most, only mildly beamed.

\section{Subsequent Monitoring and X-ray Decline}
\begin{figure}
\centering
\begin{tabular}{cc}
     \includegraphics[height=0.35\linewidth]{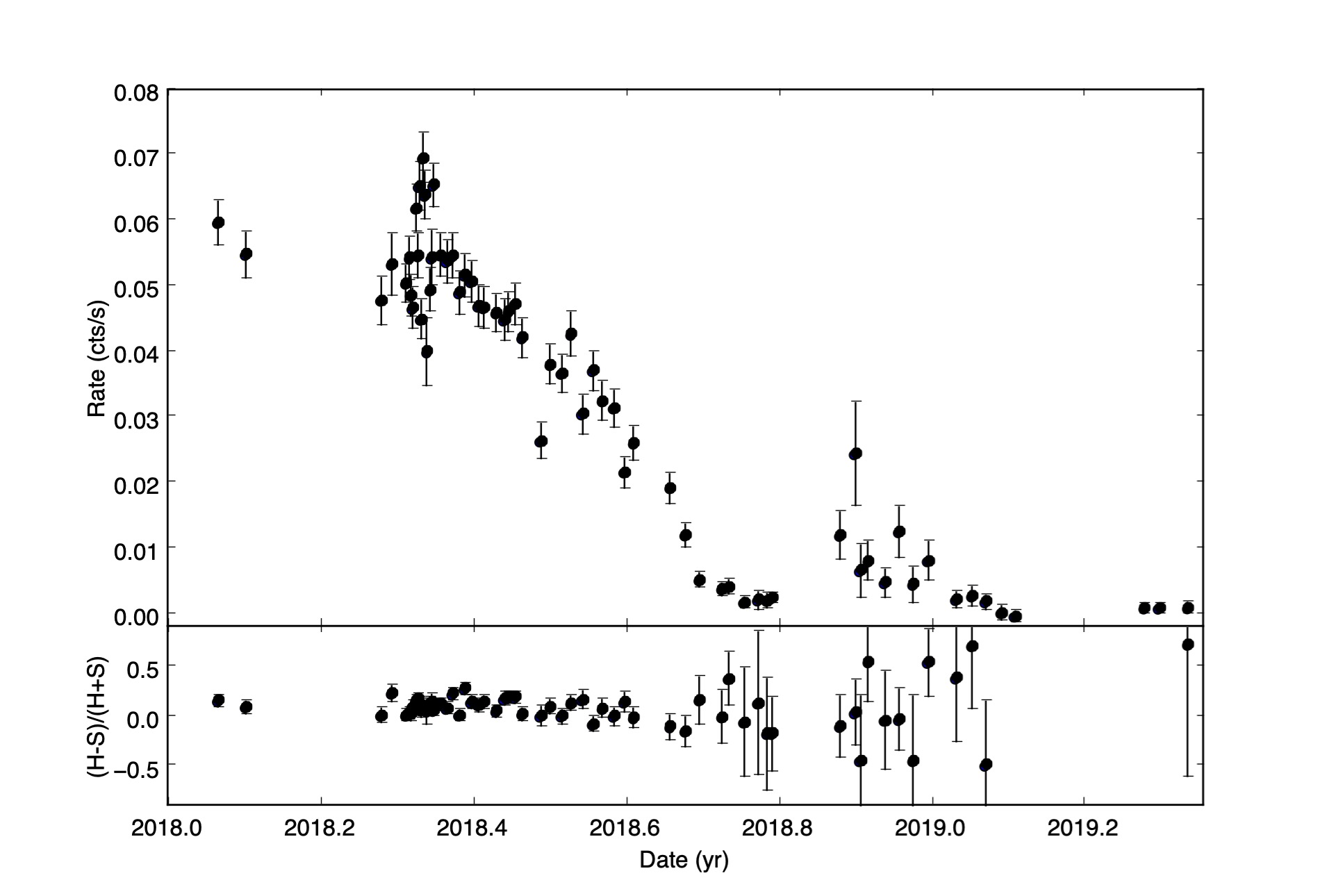} &
     \includegraphics[height=0.32\linewidth]{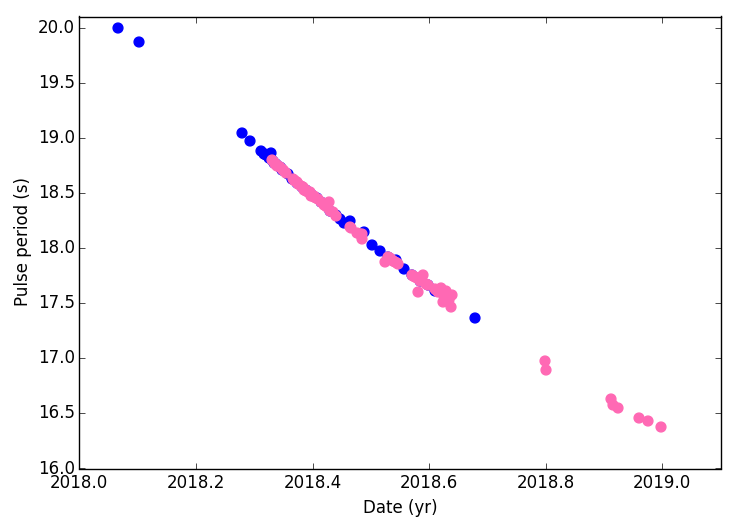} \\
\end{tabular}
\caption{{\it Left}: {\em Swift}/XRT long term light curve from 2018 January to 2019 May. The corresponding hardness ratios ($S$: 0.2-1.5\,keV, $H$: 1.5-10\,keV) are shown in the bottom panel. {\it Right}: Evolution of the pulse period from {\em NICER} (pink) and {\em Swift} (blue) data covering the period from 2018 January to 2019 January, when pulsations could be detected. }\label{figure:Swift18LC}
\end{figure}

Continual X-ray monitoring with {\em Swift} and {\em NICER}, with a few deeper {\em Chandra} and {\em NuSTAR} observations, revealed that the NGC~300 ULX-1 source stayed bright for the first half of 2018 before the flux declined and fell below {\em Swift} detectability towards the end of the year \citep{Maitra+18, Vasilopoulos+19}. The hardness ratios, however, remained nearly constant (albeit with large uncertainties), indicating the X-ray spectrum stayed constant despite the flux drop. The pulse period likewise continued to evolve as expected from earlier observations. The decrease in observed X-ray flux can be explained by an increase in obscuration near the NS, assuming a nearly constant mass accretion rate \citep[][see also Figure~\ref{figure:Swift18LC}]{Vasilopoulos+19}. At the time of writing, the last available X-ray observations of NGC~300~ULX-1 are from 2019 May. 

While some observations of NGC~300 ULX-1 can be interpreted as signatures of a clumpy outflow (e.g., the UFO detection in one set of {\em XMM-Newton} observations but not another or possible changes in partial covering fractions of high-column density material, \citep{Kosec+18,Binder+18}), the significant and prolonged decrease in X-ray flux is speculated to be the result of the Lense-Thirring effect \citep{Vasilopoulos+19,Bardeen+75,Truemper+86}, in which the accretion disk around a compact object precesses. The observed flux change is then explained as changes in the geometrical configuration of the inner portion of the accretion disk and outflow, where an observer will only see the pulsar when the wind-free region of the outflow is pointed directly along the observer's line of sight (this could also explain changes in the optical line profiles observed by \citet{Villar+16} and \citet{Binder+18}). This scenario predicts a precession period on the order of a year or longer, although the exact precession period depends sensitively on the mass accretion rate at the spherization radius (e.g., where the disk thickness becomes comparable to its radius; \citep{SS73}). Continued monitoring of NGC~300 ULX-1 for additional flux variations are needed to confirm the Lense-Thirring precession scenario for the system.

\section{The Circumstellar Environment and Donor Star}
\begin{figure}[t!]
    \includegraphics[width=0.48\linewidth]{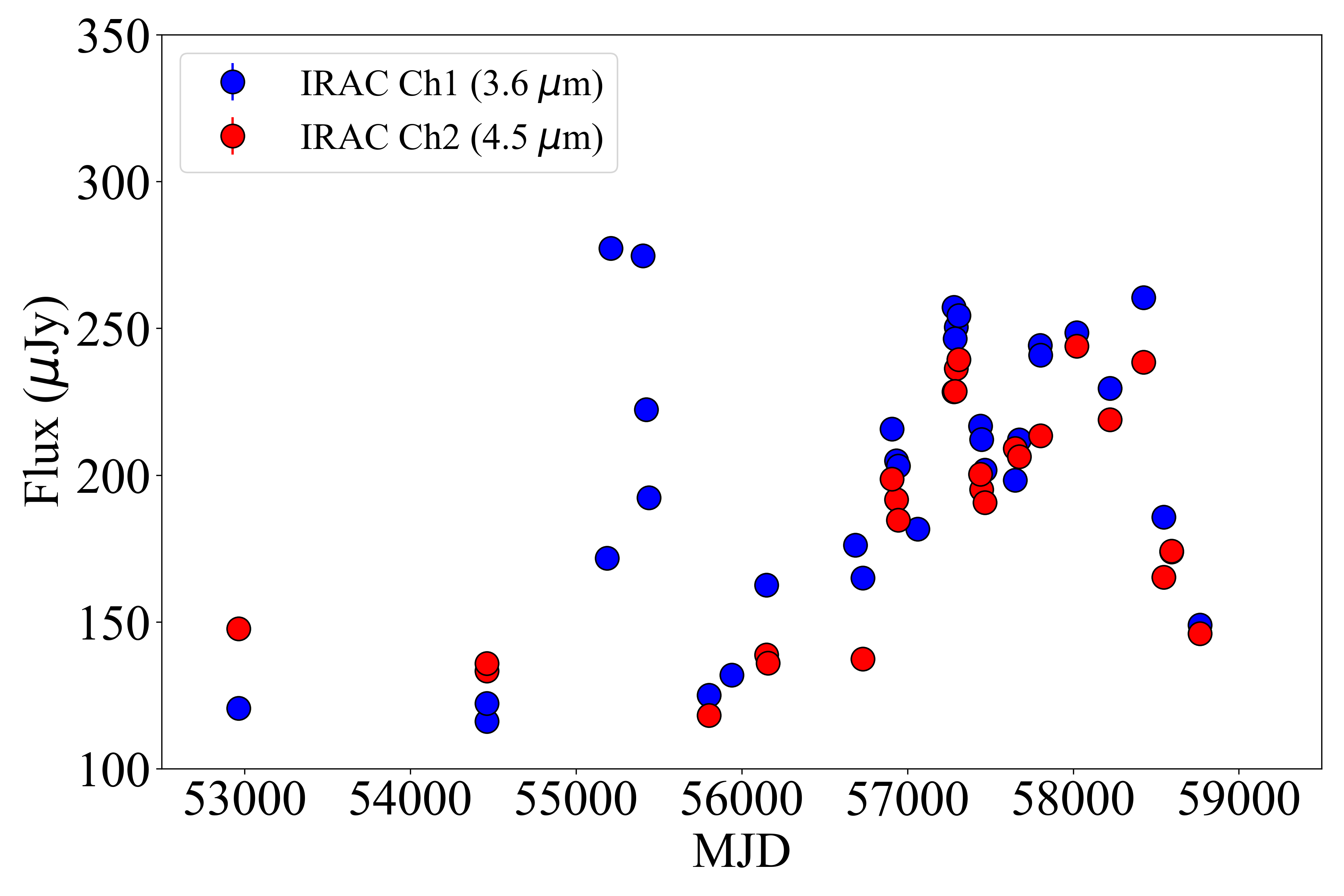} \hspace{10pt}
    \includegraphics[width=0.48\linewidth]{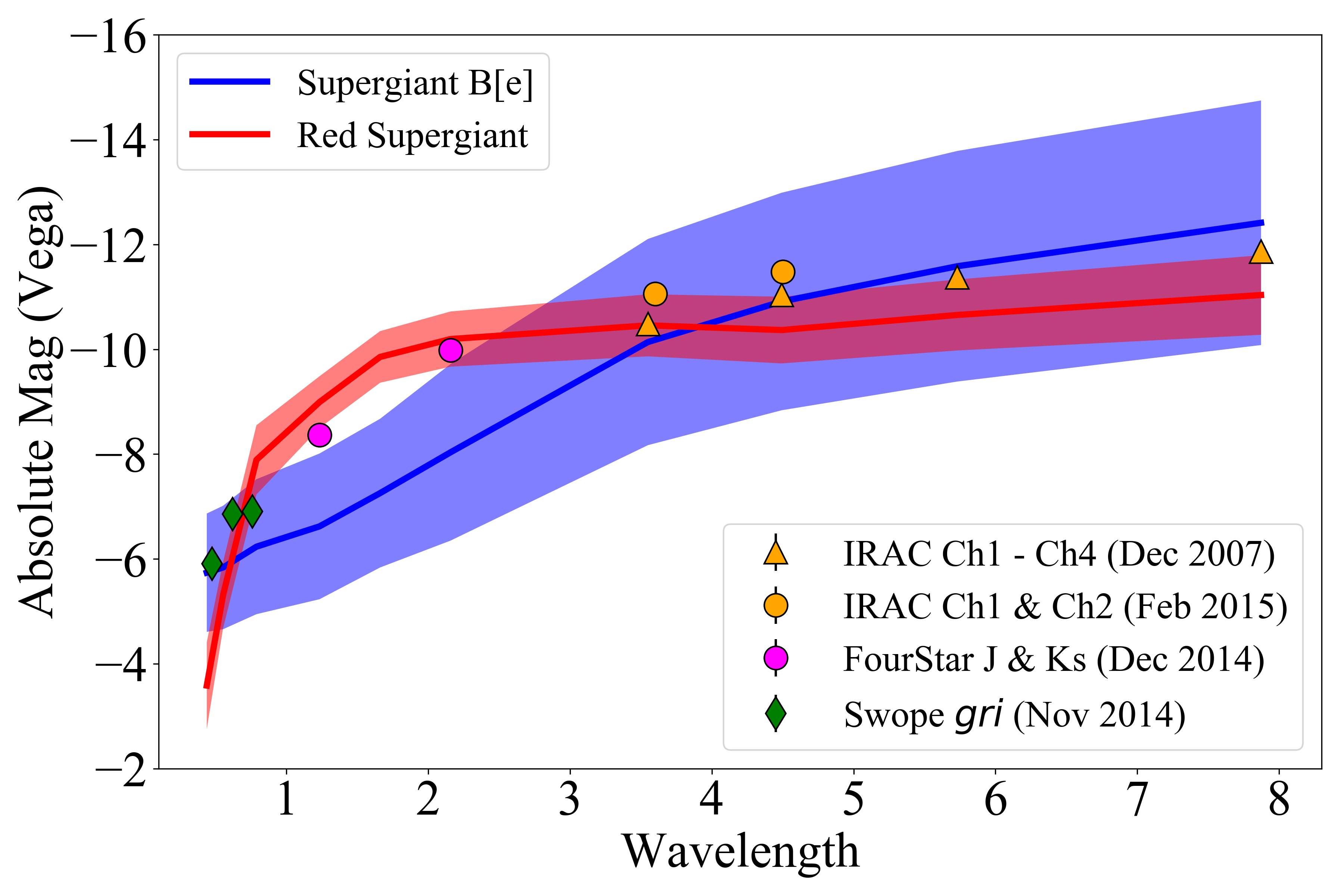}
    \caption{(\textit{Left}) \spitzer/IRAC mid-IR light curve of SN~2010da/NGC~300 ULX-1 from 2004--2019. The red and blue points correspond to photometry measured at 3.6 $\mu$m and 4.5 $\mu$m, respectively. Only 3.6 $\mu$m observations were serendipitously obtained near the 2010 outburst. (\textit{Right}) Spectral energy distribution (SED) of SN~2010da/NGC~300 ULX-1 during pre-outburst (Dec. 2007) and post-outburst phases (late 2014/early 2015) overlaid on SED templates of 122 RSG and 11 sgB[e] stars in the LMC cataloged by \citet{Bonanos+09}. Solid lines correspond to the median VIJHKs and IRAC of the SED template stars and the surrounding shaded regions indicate the 1$\sigma$ spread in the magnitudes of the distribution. The wavelength on the $x$-axis is shown in units of $\mu$m. Both figures are reproduced and modified from \citet{Lau+19}.}
    \label{fig:IR}
\end{figure}

The evolution of the emission from the infrared counterpart of NGC~300 ULX-1 presents an equally unusual mystery as the X-ray light curve. Figure~\ref{fig:IR} shows the mid-IR light curve obtained by \spitzer/IRAC over 16 years between 2004--2019 \citep{Villar+16,Lau+16, Lau+19}, where the mid-IR observations since 2014 were largely obtained as a part of the {\em Spitzer} Infrared Intensive Transients Survey (SPIRITS; \citealt{Kasliwal+17}). The mid-IR light curve exhibits three distinct phases: quiescence before the 2010 outburst, the 2010 outburst, and post-outburst variability. During the pre-outburst quiescence phase, the 4.5 $\mu$m emission exceeded the 3.6 $\mu$m emission which implied the presence of a dense circumstellar dust. In quiescence, the mid-IR 3.6--8.0 $\mu$m flux measured by \spitzer/IRAC resembled a supergiant B[e] star or a reddened RSG (Figure~\ref{fig:IR}, \citealt{Lau+19}). 

IRAC 3.6 $\mu$m observations captured the rise and decline of the 2010 outburst, which was followed by a gradual rise in mid-IR emission with intermittent flux spikes until appearing to fade significant as of 2019 March 3 (MJD 58546). The post-outburst mid-IR flux rose to a factor of $\sim2$ times higher than the progenitor but has recently faded down to the progenitor flux levels in the most recent observation taken in 2019 October. Notably, after the 2010 outburst NGC~300 ULX-1 exhibited \textit{bluer} mid-IR colors than the progenitor. In other words, in the post-outburst phase the 3.6 $\mu$m flux exceeds the 4.5 $\mu$m flux which suggests a hotter IR dust component relative to the temperatures of circumstellar dust during the quiescent progenitor phase. As shown in Figure~\ref{fig:IR}, the post-outburst optical to mid-IR SED resembles that of an RSG but is also consistent with a supergiant B[e] star. The mid-IR counterpart is slightly redder than a RSG and slightly bluer than a supergiant B[e] star, which implies that the system may host an RSG with enhanced circumstellar dust or it may be exhibiting the ``B[e]-phenomenon" with hotter circumstellar dust \citep{Lamers+98}. 

\begin{figure}
    \centering
    \includegraphics[width=0.95\linewidth]{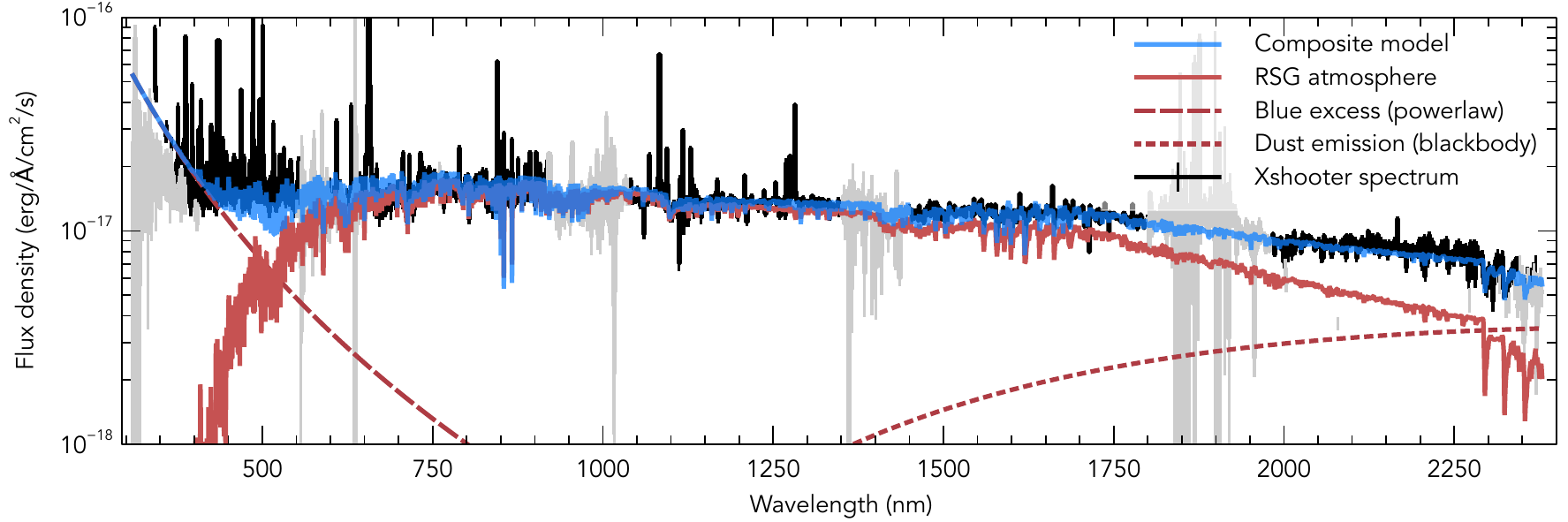}
    \caption{Deep Xshooter spectroscopy of NGC 300 ULX-1 in October 2018 (black). The composite model (blue) requires a RSG atmosphere (red), excess dust emission (red short-dashed line), and a power-law blue excess (red long-dashed line) attributed to an irradiated accretion disk. Reproduced with permission from \citet{Heida19}.}
    \label{figure:RSG_spectrum}
\end{figure}

Deep X-shooter spectroscopy covering the wavelength range from 350-2300 nm obtained in October 2018 \citep{Heida19} revealed the presence of a likely red supergiant donor star, with an effective temperature of 3650-3900 K and a bolometric luminosity $\log(L/L_{\odot}) = 4.25 \pm 0.10$. This makes the system a potential Thorne-\.{Z}ytkow object progenitor \citep{Wang16}. Two additional components are necessary to fit the X-shooter spectrum: a blackbody with a temperature $\sim 1100$ K, consistent with the presence of warm dust that was also inferred from {\em Spitzer} observations \citep{Lau+16}, and a blue component that is possibly due to emission from the irradiated accretion disk (see Figure~\ref{figure:RSG_spectrum}). The X-ray-to-optical flux ratio of ULXs generally resembles that of X-ray binaries where the optical emission is dominated by the accretion disk or X-ray heating of the stellar photosphere \citep{Kaaret+17,Tao+11}. The high degree of optical variability (in both color and magnitude) further support the hypothesis that this optical light is the result of reprocessing rather than intrinsic emission from the donor star \citep{Tao+11}. However, most X-ray binaries for which these X-ray-to-optical flux ratios have been measured contain donor stars with denser photospheres than a RSG \citep{Levesque17}, and the degree to which X-ray irradiation from an accretion disk will affect the structure of a low-density RSG photosphere remains an open question.

If accretion happens through Roche-lobe overflow, the orbital period of the system is at least 0.8-2.1 years, consistent with the lower limit of one year from X-ray timing \citep{Ray+19}. If the orbit is eccentric or the neutron star is accreting from the stellar wind or via the ``wind-Roche lobe overflow'' mechanism proposed by \citet{ElMellah+19}, in which a slow-moving wind from the donor star is beamed towards the accretor, the orbital period could be even longer. The complex structure of the Balmer lines in both the X-shooter spectrum \citep{Heida19} and spectra obtained shortly after the 2010 outburst \citep{Villar+16} both show evidence of a narrow (full-width at half-maximum of $\sim$10 \kms) component that is consistent with a RSG wind. The He~II $\lambda$4686 emission line was also detected in the X-shooter spectrum and found to be $\sim$70 times fainter than observed with Gemini by \citet{Binder+18}. Since the He~II emission feature is sensitive to the total X-ray luminosity, the change in the He~II line flux suggests ``either an overall increase of obscuring material, possibly due to increased outflow from the accretion disk, or an intrinsic drop in the X-ray luminosity,'' and not an apparent change in flux due to Lense-Thirring precession \citep{Heida19}. Additional, simultaneous X-ray and optical spectroscopic observations are needed to evaluate the degree to which the He~II $\lambda$4686 emission line and X-ray luminosity and spectral shape co-evolve.

\begin{figure}[t!]
\centering
    \includegraphics[width=0.7\linewidth]{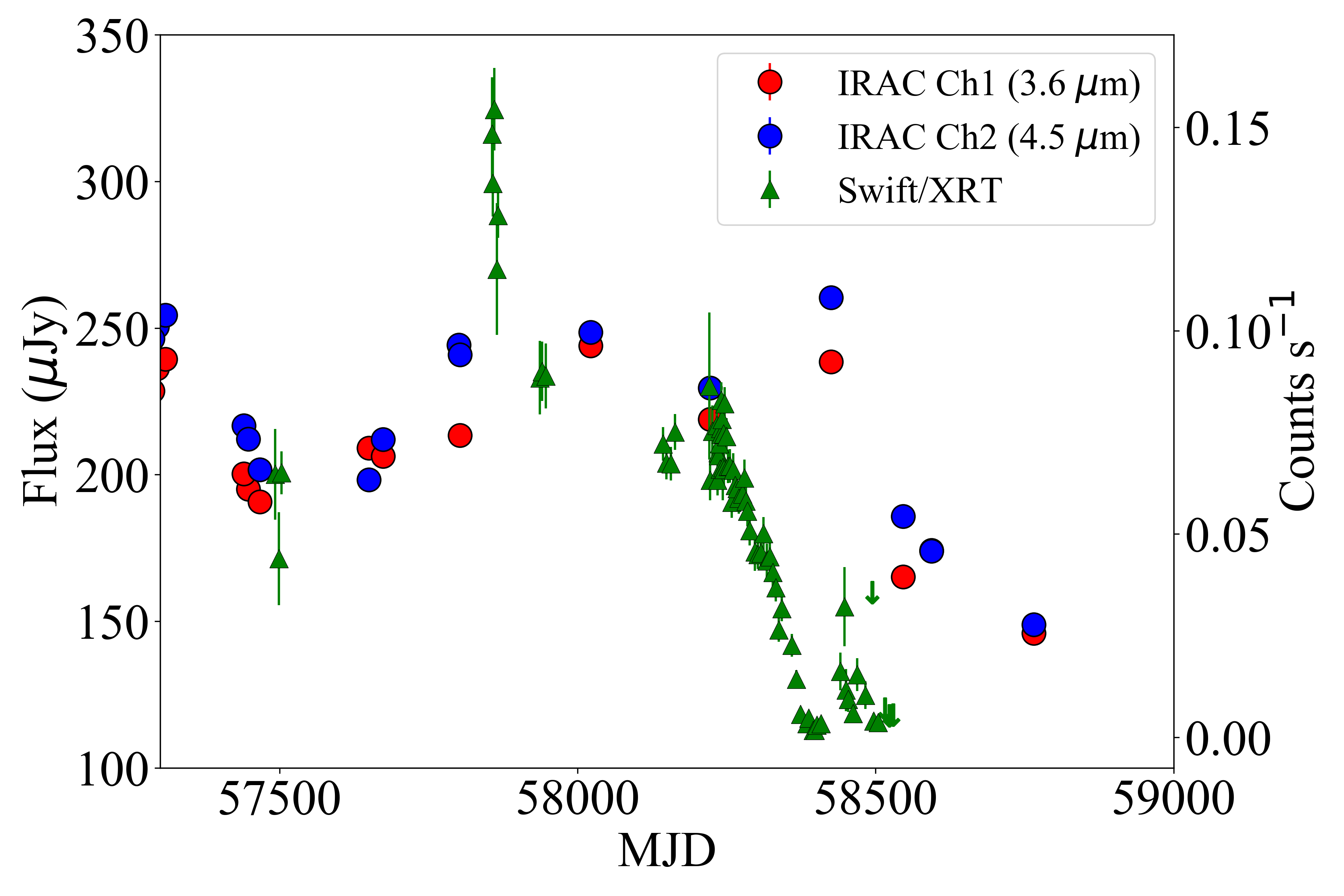} 
    \caption{Mid-IR and X-ray light curve of SN~2010da/NGC~300 ULX-1 taken between MJD 57300 and MJD 58770 by \spitzer/IRAC and \textit{Swift}/XRT, respectively. This figure is modified from \citet{Lau+19}.}
    \label{fig:IR2}
\end{figure}

\citet{Lau+19} presented the most recent mid-IR and X-ray light curve of NGC~300 ULX-1, which likely corresponds to emission from circumstellar dust and the ULX, respectively. Figure~\ref{fig:IR2} shows that mid-IR and X-ray emission taken by \spitzer\ and \textit{Swift} observations taken between 2016 and mid 2018 may be correlated. In addition to the local mid-IR/X-ray peak around 2017 April (MJD $\sim57850$), both the X-ray and mid-IR emission have decreased significantly within the past two years which suggests that the X-ray emission could contribute to heating the surrounding circumstellar dust. 

\section{Open Questions}
NGC~300 ULX-1 is, as of the time of this writing, the only ULX associated with an intermediate luminosity optical transient (or ``supernova impostor''), and the only such transient event that originated in an X-ray binary. Due to its relatively close proximity ($\sim$2 Mpc), this system offers a unique case study of a ULX containing a supergiant donor star. Despite the numerous insights into accretion mechanisms, neutron star spin evolution, and mass loss from massive stars that this system provides, several key questions about NGC~300 ULX-1 (and other ULX pulsars) remain:

\begin{itemize}
    \item What triggered the 2010 outburst? Was this event related to a common envelope phase, tidal interactions between the accretor and donor, or a change in radius of the RSG donor?
    \item What was the mass loss history of the donor star? At what rate is dust currently reforming in the binary? 
    \item What is the origin of the complex optical emission line profiles, and what can the optical spectra tell us about the orbital parameters of the binary or the geometry of the immediate circumstellar environment?
    \item Can future outbursts or ULX behavior be predicted? Are the apparent changes in X-ray flux due to Lense-Thirring precession, changes in the ultrafast outflow of the system (e.g., clumpy winds), or an intrinsic change in X-ray luminosity?
\end{itemize}

\noindent Continued photometric and spectroscopic monitoring of the source across multiple wavelength regimes is the key to answering these questions.

\authorcontributions{“Conceptualization, B.A.B.; writing—original draft preparation, B.A.B., S.C., M. H., R. L.; writing—review and editing, visualization, B.A.B.}

\funding{This research received no external funding.}

\acknowledgments{The authors would like to thank the anonymous referees for comments and suggestions that significantly improved this manuscript. The authors} would also like to thank Edo Berger and Rubab Khan for granting permission to reproduce figures from {\it The Astronomer's Telegram} postings, and Emily Levesque and Benjamin Williams for comments that improved early versions of this manuscript.

\conflictsofinterest{The authors declare no conflict of interest.}

\reftitle{References}

\end{document}